\documentclass[conference]{IEEEtran}
\usepackage[backend=biber,style=ieee]{biblatex}
\usepackage{amsmath}
\usepackage{graphicx}
\usepackage{multirow}
\usepackage{threeparttable}

\usepackage{flushend}
\usepackage{amsmath,amssymb,amsfonts}
\usepackage{algorithmic}
\usepackage{graphicx}
\usepackage{textcomp}
\usepackage{xcolor}
\usepackage{booktabs}
\usepackage{array}

\usepackage{geometry}
\usepackage{fancyhdr}

\fancypagestyle{firststyle}{
  \fancyhf{}
  \fancyhead[C]{2025 Asia Pacific Signal and Information Processing Association Annual Summit and Conference (APSIPA ASC)}
}

\begin{document}

\title{Multimodal Sentiment Analysis with Missing Modality: A Knowledge-Transfer Approach}

\iffalse
\author{
\authorblockN{
Weide Liu\authorrefmark{1} and Huijing Zhan\authorrefmark{2}
}

\authorblockA{
\authorrefmark{1} College of Computing and Data Science, Nanyang Technological University, Singapore \\
E-mail: weide001@e.ntu.edu.sg}

\authorblockA{
\authorrefmark{2}
Singapore University of Social Sciences, Singapore\\
E-mail: hjzhan@suss.edu.sg}
}
\fi 
\author{
\IEEEauthorblockN{Weide Liu and Huijing Zhan}
\IEEEauthorblockA{
College of Computing and Data Science, Nanyang Technological University, Singapore\\
Singapore University of Social Sciences, Singapore\\
Email: weide001@e.ntu.edu.sg, hjzhan@suss.edu.sg
}
}
\maketitle
\thispagestyle{firststyle}
\pagestyle{fancy}

\begin{abstract}
Multimodal sentiment analysis aims to identify the emotions expressed by individuals through visual, language, and acoustic cues. However, most existing research assume that all modalities are available during both training and testing, which makes their algorithms susceptible to the missing-modality scenarios. In this paper, we propose a novel knowledge-transfer network to translate between different modalities to reconstruct the missing audio features. Moreover, we develop a cross-modality attention mechanism to maximize the information extracted from the reconstructed and observed modalities for sentiment prediction. Extensive experiments on three publicly available datasets demonstrate significant improvements over baseline methods and achieve comparable results to the previous methods with complete multi-modality supervision.

\end{abstract}

\section{Introduction}
\label{sec:intro}
With the rapid advancement of deep learning~\cite{liu2020crnet,liu2025modality}, multimodal sentiment analysis (MSA) has emerged as a growing area of research in recent years, as it allows for a more comprehensive and effective understanding of an individual's emotions. Multiple sources of information are utilized, including visual, language, and acoustic modalities, which together provide a more complete picture of an individual's emotional state.

Recent works on MSA~\cite{tsai2019multimodal,cmu-mosi,DBLP:conf/cvpr/ZhangGWZLCCRHYC16,DBLP:conf/cvpr/ZengTPLZZHL05,DBLP:conf/iccv/GanWHJ17,transformer,chen2018best,lee2018pre,liang2018multimodal,wang2019words} have focused on developing effective methods for integrating and utilizing multi-modality information, under the assumption that all modalities are available during both training and testing. However, in real-world scenarios, missing modality is a common problem due to privacy concerns or technical difficulties. Especially in situations such as online meetings and network sharing, where data is frequently uploaded and downloaded, modalities can be missing during transmission. In these cases, it becomes essential to reconstruct missing modalities using the information from observed modalities.

Previous research studies \cite{pham2019found,tsai2018learning} have attempted to address the issue of missing modalities in multimodal sentiment analysis. In particular, Tsai \textit{et al.} \cite{tsai2018learning} proposed a joint generative-discriminative objective to obtain a robust multimodal representation and a surrogate inference model for missing modalities. Pham \textit{et al.} \cite{pham2019found} developed a multimodal translation network with a cyclic translation loss for forward adaptation between source and target modalities. However, the performance of their approaches degrades when complete modality information is not available during the training stage. 

In this work, we propose a knowledge-transfer network to reconstruct missing acoustic modalities using transformer blocks and a consistency loss as a constraint during training. Additionally, we introduce a cross-modal attention network to effectively fuse representations from available modalities and the reconstructed features for a robust joint multimodal representation. This allows for more informative signals to be emphasized in the cross-modal attention blocks, leading to improved multimodal representation learning.
Experiments on three multimodality sentiment analysis datasets indicate that our method can achieve comparable performance to those using complete modality supervision.

\begin{figure*}[t]
  \centering
    \includegraphics[width=1\linewidth]{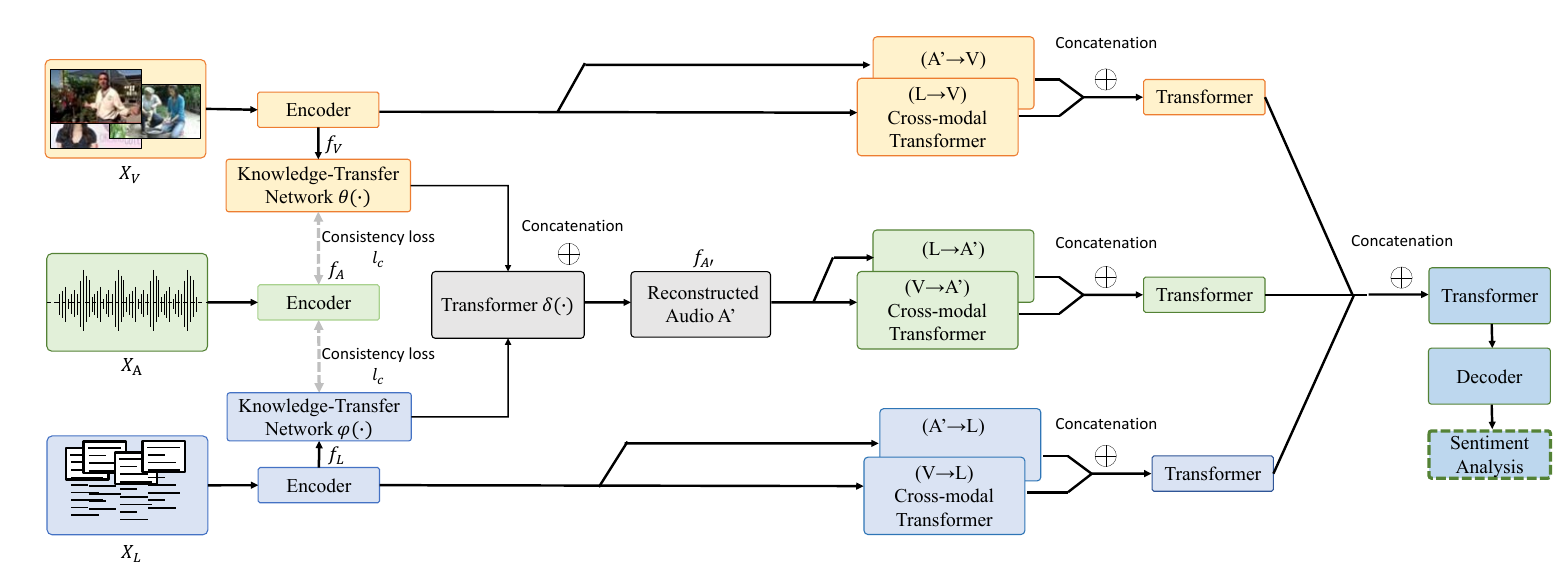}
    \caption{The pipeline of our method. The A' denotes the reconstructed audio information.}
    \label{Figure: pipeline}
    \vspace{-2mm}
\end{figure*}

The main contributions of this work are concluded from those aspects:
\begin{itemize}
    \item To the best of our knowledge, we are among the pioneering work to address the problem of missing modality imputation with the transformer framework.
    \item We propose a novel knowledge-transfer network to reconstruct the missing modality from available modalities.
    \item Extensive experiments validate the effectiveness of our proposed method on multimodal sentiment analysis and its robustness in the missing modality scenarios.
\end{itemize}

\section{Proposed Method}
As illustrated in Fig.~\ref{Figure: pipeline}, during training, given a video with visual, language, and acoustic modalities, denoted as $X_v$, $X_L$, and $X_A$, the modality-specific encoder independently maps each modality into its modality-aware feature. As we are addressing the problem of missing modality, we assume that the acoustic feature is not directly involved in multi-modality representation learning. More specifically, we propose a novel knowledge-transfer network to reconstruct the acoustic feature based on the available visual or language signals. The mutual relationships between the visual, language, and reconstructed acoustic features are modeled using the cross-modal attention network, which consists of a set of transformer blocks. Finally, the aggregated features are then used for the multimodal human language sentiment analysis.

\subsection{Knowledge-Transfer Network}
In this section, we will present our approach for reconstructing the missing audio modality information from the available visual and language modality features. We use a set of transformer-based encoders to convert the vision, language, and audio modalities into modality-specific features, denoted as $\mathbf{f}_V$, $\mathbf{f}_L$, and $\mathbf{f}_A$ respectively. To ensure the reconstructed audio feature is close to the ground truth, we employ the consistency loss function $\ell_c$, which minimize the Euclidean distance between the reconstructed and ground truth audio features, as defined below:

\begin{equation}
    \mathcal{L}_{V \rightarrow A}^c = \ell_c (\theta(\mathbf{f}_V) , \mathbf{f}_A), \quad
    \mathcal{L}_{L \rightarrow A}^c = \ell_c (\phi (\mathbf{f}_L) , \mathbf{f}_A),
    \label{equ:loss c}
\end{equation}
where the $\theta$ and the $\phi$ stand for the knowledge-transfer network for visual and language modality, consisting of a set of transformer blocks. Here $\theta(\mathbf{f}_V)$ and $\phi (\mathbf{f}_L)$ denote the acoustic features reconstructed from the visual and the language modality, respectively. %In this paper, we have tried two kinds of consistency loss functions, including L1 loss and L2 loss. 

In order to reconstruct as much of the missing audio modality as possible, multiple available modalities are leveraged as supervision to train our network. Specifically, we combine the reconstructed acoustic features $\mathbf{f}_A'$ from both the visual and language modalities, as shown below:

\begin{equation}
    \mathbf{f}_A^{'} = \delta ([\theta(\mathbf{f}_V) ||  \phi(\mathbf{f}_L)]),
    \label{equ:concate}
\end{equation}
where $||$ denotes the concatenation operation which combines the reconstructed acoustic features from vision $\theta(\mathbf{f}_V)$ and language $\phi(\mathbf{f}_L)$. Instead of intuitive concatenation, we utilize a set of transformer blocks $\delta$ to encode the combined reconstructed acoustic information for effective representation.

\subsection{Cross-modal Attention}
To obtain a comprehensive joint multi-modal representation, it is essential to model the inter-dependency relationship between different modalities. Inspired by MulT~\cite{tsai2019multimodal}, we also consider fusing cross-modal information by providing a latent adaptation across modalities. Consider a source modal feature $m$ and a target modal feature $m'$, we map the target modal features $m'$ into a latent space as the query $Q_{m'} = m' W_{Q_{m'}}$, the key and value are obtained from the source modal features $K_m = m W_{K_m}$, and $V_m = m W_{V_m}$, where $W_{Q_{m'}}, W_{K_m}$ and $W_{V_m}$ are the weights. The latent adaptation from the source modality $m$ to the target query feature $m'$ is presented as the cross-modal attention (CM), $Z_m' := \text{CM}_{m \rightarrow m'}(m', m)$:

\vspace{2mm}
\begin{equation}
\label{eq:cross-modalBA}
\resizebox{0.7\hsize}{!}{
$
\begin{aligned}
    Z_m' &= \text{CM}_{m \rightarrow m'}(m', m) \\
    &= \text{softmax}\left(\frac{Q_{m'} K_m^\top}{\sqrt{d_k}}\right) V_m
    \\
    &= \text{softmax}\left(\frac{m' W_{Q_m'} W_{K_m}^\top m^\top}{\sqrt{d_k}}\right) m W_{V_m},
\end{aligned}
$}
\vspace{-3mm}
\end{equation}
where $\sqrt{d_k}$ denotes the scaled parameter, and the ${d_k}$ denotes the length of the target features.
%Following the previous transformer works~\cite{transformer,chen2018best,lee2018pre,dai2018transformer}, 
A residual connection is utilized to connect the original target feature $m'$ and attended features $Z_m'$ after the cross-modal attention computation as our final feature:
\begin{equation}
    m' = m' + Z_m'.
\end{equation}
Finally, the attended target feature with inter-modality correlation information are subsequently combined for the sentiment analysis prediction task, with the prediction denoted as $y'$. Note that the decoder consists of a set of self-attention transformer~\cite{transformer}. The standard cross entropy loss function $\ell_{ce}$ is utilized to compute the difference between the ground truth $y$ and the prediction $y'$:

\begin{equation}
\mathcal{L}_{e} = \ell_{ce}(y,y'). 
\end{equation}
 The overall loss function $L$ to optimize is defined as below: 
   \begin{equation}
 \mathcal{L}=\mathcal{L}_{e}+\lambda_1 \mathcal{L}_{V \rightarrow A}^{c} + \lambda_2 \mathcal{L}_{L \rightarrow A}^{c},
 \label{loss_sum}
  \end{equation}
where $\lambda_1$ and $\lambda_2$ refer to the trade-off parameters for the consistency loss from the visual and language modality, respectively.

\section{Experiment}
% Following previous methods~\cite{pham2019found, tsai2018learning, liang2018multimodal}, we conduct experiments on three benchmark datasets for sentiment analysis and emotion recognition to validate our methods.

Following the previous methods~\cite{pham2019found, tsai2018learning, liang2018multimodal}, we conducted experiments on three benchmark datasets for sentiment analysis and emotion recognition to validate the effectiveness of our proposed method.

\begin{table}[t!]
\caption{\small Results for multimodal sentiment analysis on the CMU-MOSI dataset with aligned and non-aligned multi-modal sequences.  $^\uparrow$: the higher, the better. $^\downarrow$: the lower, the better.
}
\begin{center}
\fontsize{9}{10}\selectfont
\setlength\tabcolsep{2pt}
\begin{tabular}{|l||c|c|c|c|c|}
\hline
Metric       & Acc$_7^\uparrow$  & Acc$_2^\uparrow$  & F1$^\uparrow$    & MAE$^\downarrow$   & Corr$^\uparrow$  \\ \hline \hline
\multicolumn{6}{|c|}{(Word Aligned) CMU-MOSI Sentiment}     \\ \hline
RMFN~\cite{liang2018multimodal}     & 38.3  & 78.4  & 78.0  & 0.922 & 0.681 \\
MFM~\cite{tsai2018learning}       & 36.2  & 78.1  & 78.1  & 0.951 & 0.662 \\
RAVEN~\cite{wang2019words}          & 33.2     & 78.0  & 76.6    & 0.915 & 0.691 \\
MCTN~\cite{pham2019found}          & 35.6     & 79.3  & 79.1  & 0.909 & 0.676 \\ \hline
Full modality (ours)    & 39.7  & 82.9  & 82.7  & 0.870 & 0.694 \\ \hline
Vision only (ours)      & 34.1  & 76.6  & 76.4  & 0.721 & 0.421 \\ \hline
Language only (ours)    & 36.4  & 78.1  & 77.3  & 0.751 & 0.619 \\ \hline
Language and vision (ours) & 38.7  & 81.3  & 81.2  & 0.849 & 0.688 \\ \hline
Ours & {39.1}  & {82.3}  & {82.2}  & {0.858} & {0.691} \\ \hline
\hline
\multicolumn{6}{|c|}{(Unaligned) CMU-MOSI Sentiment} \\ \hline
CTC~\cite{ctc} + MCTN~\cite{pham2019found}     & 32.7     & 75.9  & 76.4  & 0.991 & 0.613 \\
CTC~\cite{ctc} + RAVEN~\cite{wang2019words}    & 31.7     & 72.7  & 73.1  & 1.076 & 0.544 \\ \hline
Full modality (ours)    & 39.3  & 82.3  & 82.1  & 0.861 & 0.690 \\ \hline 
Vision only(ours)       & 34.0  & 76.1  & 75.9  & 0.717 & 0.401 \\ \hline
Language only (ours)    & 36.0  & 77.5  & 77.2  & 0.742 & 0.606 \\ \hline
Language and vision (ours) & 37.5  & 80.9  & 80.7  & 0.838 & 0.681 \\ \hline
Ours &38.7  &81.9  & 81.7  & 0.844 & 0.687 \\ \hline
\end{tabular}
\end{center}
\label{table:mosi}
\end{table}

\subsection{Datasets and Experimental Settings}

\noindent\textbf{CMU-MOSI \& MOSEI.} CMU-MOSI~\cite{cmu-mosi} is a multimodal sentiment analysis dataset containing 2,199 short monologue video clips. CMU-MOSEI~\cite{cmu-mosei} consists of 23,454 video clips from YouTube, and each sample is assigned a sentiment score by human annotators, ranging from -3 (strongly negative) to 3 (strongly positive). Following previous methods~\cite{tsai2019multimodal, pham2019found, zadeh2018memory}, the performances are measured with a variety of evaluation metrics, including 7-class sentiment score classification (Acc$_7$), binary positive/negative sentiments prediction accuracy (Acc$_2$), F1 score, mean absolute error (MAE), and correlation of the model's prediction with subjective study (Corr).

\noindent\textbf{IEMOCAP}
IEMOCAP~\cite{iemocap} is a multi-label emotion recognition dataset that contains around 10,000 videos. The dataset includes four classes: happy, sad, angry, and neutral. Unlike CMU-MOSI~\cite{cmu-mosi} and CMU-MOSEI~\cite{cmu-mosei}, this dataset focuses on multi-label prediction, where a person can express multiple emotions simultaneously. Following the previous methods~\cite{tsai2019multimodal, poria2017context, wang2019words}, the binary classification accuracy (Acc) and F1 score are reported in the experiments.

\noindent\textbf{Implementation Details}
In this paper, we utilize the Multimodal Transformer~\cite{tsai2019multimodal} as our backbone and baseline. The transformer blocks within the model consist of three transformer layers~\cite{transformer}. We employ Adam optimization with 40 training epochs and maintain a constant learning rate of 1e-3 throughout the training process. We utilize the same training and testing split as that in \cite{tsai2019multimodal} and CTC \cite{ctc} is applied on the unaligned setting of baseline approaches.

\subsection{Performance Comparison}
We compare our results to state-of-the-art methods that utilize full modalities for supervision. Our method, however, is evaluated with certain missing modalities. The ``Vision only'' denotes that only the visual modal information (yellow sub-branch in Fig.~1) is utilized as input. ``Language only'' denotes that only the language modal information (blue sub-branch in Fig.~1) is utilized for training the model. ``Language and vision'' denotes that we fuse the visual and language modal information with the proposed cross-modal fusion module, but without the reconstructed acoustic modal information from our knowledge-transfer network. ``Full modality'' denotes the complete modality supervision.

\noindent\textbf{CMU-MOSI.}
As shown in Table~\ref{table:mosi}, our method outperforms the lower bound results (including vision and language) for both the aligned and unaligned datasets. It's worth noting that, with single-modal information, language performance is better than vision, indicating that language is more important than vision modal information in this dataset. When fusing vision and language without our reconstructed audio feature, the performance is significantly improved over methods with only one modal information, which demonstrates that more modal information can improve the performance. Our method achieves competitive results compared to the upper bound with the use of our reconstructed audio features. These results indicate that our reconstructed feature can effectively restore the missing audio modal information.

\newcolumntype{K}[1]{>{\centering\arraybackslash}p{#1}}
\begin{table}[t!]
%\vspace{-.1mm}
\caption{\small Results for multimodal sentiment analysis on CMU-MOSEI dataset with aligned and non-aligned multimodal sequences. $^\uparrow$: the higher, the better. $^\downarrow$: the lower, the better.}
\begin{center}
\fontsize{8}{10}\selectfont
%\centering
\setlength\tabcolsep{0.7pt}
%\vspace{-4mm}
\begin{tabular}{|c||*{5}{K{0.72cm}}|}
\hline
{Metric}         & {Acc$_7^\uparrow$}  & {Acc$_2^\uparrow$}  & {F1$^\uparrow$}    & {MAE$^\downarrow$}   & {Corr$^\uparrow$}  \\ \hline \hline
\multicolumn{6}{|c|}{{(Word Aligned) CMU-MOSEI Sentiment}}     \\ \hline \hline
% EF-LSTM        & 47.4  & 78.2  & 77.9  & 0.642 & 0.616 \\
% LF-LSTM        & 48.8    & 80.6  & 80.6  & 0.619 & 0.659 \\
Graph-MFN~\cite{zadeh2018memory}           & 45.0  & 76.9  & 77.0  & 0.71 & 0.54 \\
RAVEN~\cite{wang2019words}          & 50.0     & 79.1  & 79.5     & 0.614 & 0.662 \\ 
MCTN~\cite{pham2019found}           & 49.6     & 79.8  & 80.6  & 0.609 & 0.670 \\ \hline \hline

Full modality (ours)           &       {50.7}     &       {80.6}  &       {80.8}  &       {0.623} &       {0.700} \\ \hline

Vision only (ours)          & 43.5    &  66.4  &  69.3  &        {0.756} &  0.343 \\ \hline

Language only (ours)          & 46.5    &  77.4  &  78.2  &  0.653 &  0.631 \\ \hline

Language and vision (ours)          & 48.4    &  79.5  &  79.6  & 0.639 &  0.633 \\ \hline

Ours            &        {51.1}     &        {80.0}  &        {80.3}  &  0.635 &        {0.637} \\ \hline

\hline
\multicolumn{6}{|c|}{(Unaligned) CMU-MOSEI Sentiment} \\ \hline \hline
% CTC + EF-LSTM  & 46.3  &  76.1  & 75.9  & 0.680 & 0.585   \\
% LF-LSTM        & 48.8     & 77.5  & 78.2    &  0.624 & 0.656 \\
CTC~\cite{ctc} + RAVEN~\cite{wang2019words}    & 45.5     & 75.4  & 75.7  & 0.664 & 0.599 \\ 
CTC~\cite{ctc} + MCTN~\cite{pham2019found}     & 48.2     & 79.3  & 79.7  & 0.631 &  0.645 \\ \hline \hline

Full modality (ours)           &       {49.7}     &       {79.8}  &       {80.1}  &       {0.641} &       {0.681} \\ \hline

Vision only (ours)          & 42.1    &  65.7  &  68.4  &        {0.741} &  0.339 \\ \hline

Language only (ours)          & 45.4    &  76.9  &  77.1  &  0.660 &  0.623 \\ \hline

Language and vision (ours)          & 47.9    &  78.2  &  79.1  & 0.649 &  0.627 \\ \hline

Ours            &        {49.6}     &        {79.4}  &        {79.5}  &  0.646 &        {0.648} \\ \hline

\end{tabular}
\end{center}
\label{table:mosei}
% \vspace{-2mm}
\end{table}

\noindent\textbf{CMU-MOSEI.}
As depicted in Table~\ref{table:mosei}, the results for both aligned and unaligned settings for the MOSEI dataset demonstrate that our method attains comparable performance to the fully-supervised method and surpasses all previous methods. Interestingly, our method demonstrates even better performance than the full modality supervision on the evaluation metrics of $Acc_7$ and MAE on the aligned setting. This could be due to the fact that some of the audio information provided by the ground truth audio might negatively impact the prediction, such as background noise. For example, in a video of a boy who appears sad, his feelings can be accurately reflected by his facial expressions and speech, but if the audio includes background laughter, it may mislead the network's ability to make an accurate judgement. However, our reconstructed audio modal information is obtained from vision and text, which is not affected by background noise, thus providing more accurate results.

Interestingly, our method occasionally surpasses the full-modality baseline, particularly in CMU-MOSEI. This outcome arises because real acoustic signals sometimes contain irrelevant or misleading noise (e.g., background laughter, music, or microphone distortion) that can bias the prediction. In contrast, the reconstructed audio representation is derived from visual and textual cues, effectively denoising the audio stream. As a result, our reconstructed features may be more informative for sentiment recognition than raw acoustic input, leading to performance exceeding the full-modality upper bound.

\begin{table}[t!]
\vspace{-5mm}
\caption{\small Results for multimodal emotions analysis on IEMOCAP with aligned and non-aligned multimodal sequences.
}
\begin{center}
\fontsize{8}{10}\selectfont
%\centering
\setlength\tabcolsep{0.7pt}
%\vspace{-4mm}
\begin{tabular}{|c||*{8}{K{0.53cm}}|}
\hline
Task          & \multicolumn{2}{c}{Happy} & \multicolumn{2}{c}{Sad} & \multicolumn{2}{c}{Angry} & \multicolumn{2}{c|}{Neutral} \\
Metric        & Acc$^\uparrow$         & F1$^\uparrow$          & Acc$^\uparrow$        & F1$^\uparrow$         & Acc$^\uparrow$         & F1$^\uparrow$          & Acc$^\uparrow$           & F1$^\uparrow$           \\ \hline \hline
\multicolumn{9}{|c|}{(Word Aligned) IEMOCAP Emotions}                                                                               \\ \hline  \hline
% EF-LSTM       & 86.0        & 84.2        & 80.2       & 80.5       & 85.2        & 84.5        & 67.8          & 67.1         \\
% LF-LSTM       & 85.1        & 86.3        & 78.9       & 81.7       & 84.7        & 83.0        & 67.1          & 67.6         \\
RMFN~\cite{liang2018multimodal}          & 87.5        & 85.8        & 83.8       & 82.9       & 85.1        & 84.6        & 69.5          & 69.1         \\
MFM~\cite{tsai2018learning}           & 90.2        & 85.8        &  {88.4}    &  {86.1}       &  {87.5}        & 86.7        & 72.1          & 68.1         \\
RAVEN~\cite{wang2019words}         & 87.3        & 85.8        & 83.4       & 83.1     &   {87.3}        & 86.7        & 69.7          & 69.3         \\
MCTN~\cite{pham2019found}          & 84.9          & 83.1           & 80.5        & 79.6          & 79.7          & 80.4           & 62.3            & 57.0            \\ \hline  \hline
Full modality (ours)          &      {90.3}        &      {88.1}        &     {86.4}       &      {86.0}       &      {87.3}        &      {87.0}        &      {72.2}          &      {70.1}         \\ \hline  
Vision only (ours)          &  {83.7}        &  {81.6}        & 81.5       &  {81.2}       &  {82.0}        &  {81.3}        &  {63.2}          &  {62.7}         \\ \hline  
Language only (ours)            &  {85.3}        &  {85.9}        & 85.7       &  {84.2}       &  {86.1}        &  {85.6}        &  {70.1}          &  {68.7}         \\ \hline  
Language and vision (ours)             &  {89.1}        &  {86.8}        & 85.9       &  {85.0}       &  {86.1}        &  {85.2}        &  {70.0}          &  {69.4}         \\ \hline  
Ours       &     {90.1}        &     {87.6}        &    {87.5}       &     {85.5}       &     {87.2}        &     {86.8}        &     {71.9}          &     {70.1}         \\ \hline   

\hline
\multicolumn{9}{|c|}{(Unaligned) IEMOCAP Emotions}                                                                           \\ \hline  \hline
% CTC~\cite{graves2006connectionist} + EF-LSTM  &  76.2  &  75.7  & 70.2  & 70.5  & 72.7  &  67.1  &  58.1 &  57.4 \\
% LF-LSTM       & 72.5  &  71.8  &  72.9  &  70.4   &  68.6  &  67.9  &   59.6   &   56.2     \\
CTC~\cite{ctc} + RAVEN~\cite{wang2019words}  &  77.0  &  76.8  & 67.6  &  65.6  & 65.0  & 64.1  &   {62.0}  &   {59.5} \\
CTC~\cite{ctc} + MCTN~\cite{pham2019found}   &  80.5  & 77.5  & 72.0  & 71.7 &   64.9  &  65.6  &  49.4  &  49.3 \\  \hline \hline
Full modality (ours)         &      {84.8}        &      {81.9}        &      {77.7}       &      {74.1}       &      {73.9}        &      {70.2}        &      {62.5}          &      {59.7}      \\ 

Vision only (ours)          &  {77.7}        &  {72.6}        & 69.9       &  {68.3}       &  {68.1}        &  {64.6}        &  {57.3}          &  {51.2}         \\ \hline  
Language only (ours)            &  {79.3}        &  {78.8}        & 74.6       &  {71.8}       &  {71.6}        &  {67.5}        &  {60.2}          &  {55.9}         \\ \hline  
Language and vision (ours)             &  {82.8}        &  {81.0}        & 76.9       &  {72.8}       &  {72.6}        &  {68.2}        &  {60.9}          &  {58.4}         \\ \hline  
Ours       &     {84.4}        &     {81.7}        &    {77.7}       &     {74.0}       &     {73.8}        &     {69.6}        &     {61.9}          &     {59.5}         \\ \hline   

\hline
\end{tabular}
\end{center}
\label{table:iemocap}
%\vspace{-5mm}
\end{table}

\noindent\textbf{IEMOCAP.}
As evidenced by the results presented in Table~\ref{table:iemocap}, our method demonstrates comparable performance to the full modality method and surpasses all baseline methods on both the aligned and unaligned settings. This conclusion is consistent with the results observed in the other two datasets, CMU-MOSI and CMU-MOSEI.

% \begin{table}[t]

% \centering
% \small

% \caption{The effectiveness of each cross-modal transformer. 
% The results are reported with the CMU-MOSEI aligned dataset~\cite{cmu-mosei}.}
% \vspace{-3mm}
% \resizebox{0.75\linewidth}{!}{
% \begin{tabular}{|l|c|c|c|c|c|}
% \hline
% Target Modal    & Acc$_7^\uparrow$  & Acc$_2^\uparrow$  & F1$^\uparrow$    & MAE$^\downarrow$   & Corr$^\uparrow$  \\ \hline
% Language & 49.0 & 79.7 & 80.2 & 0.636 & 0.632 \\
% Audio & 48.2 & 79.6 & 80.0 & 0.639 & 0.627 \\
% Vision & 48.4 & 79.5 & 79.6 & 0.641 & 0.633 \\ \hline
% Ours            &    {51.1}     &   {80.0}  &    {80.3}  &    {0.635} &    {0.637} \\
% \bottomrule

% \end{tabular}
% }
% \label{Table:abalation}
% \end{table}

% \newcolumntype{K}[1]{>{\centering\arraybackslash}p{#1}}
\begin{table}[t]
\centering
\small
\caption{\small The effectiveness of each cross-modal transformer. 
The results are reported on the CMU-MOSEI aligned dataset~\cite{cmu-mosei}.}
\resizebox{0.9\linewidth}{!}
{
\begin{tabular}{l|c|c|c|c|c}
\toprule
Target Modal & Acc$_7^\uparrow$ & Acc$_2^\uparrow$ & F1$^\uparrow$ & MAE$^\downarrow$ & Corr$^\uparrow$ \\ \midrule
%Target Modal & Acc & Acc & F1 & MAE & Corr \\ \midrule
Language & 49.0 & 79.7 & 80.2 & 0.636 & 0.632 \\
Audio & 48.2 & 79.6 & 80.0 & 0.639 & 0.627 \\
Vision & 48.4 & 79.5 & 79.6 & 0.641 & 0.633 \\
Ours &          {51.1} &          {80.0} &          {80.3} &          {0.635} &          {0.630} \\
\bottomrule

\end{tabular}
\label{Table:abalation}
}
\end{table}

% \begin{table}[t]

% \centering
% \small

% \caption{The effectiveness of different consistency Loss. The results are reported with the CMU-MOSEI aligned dataset~\cite{cmu-mosei}.}
% \vspace{-3mm}
% \resizebox{0.7\linewidth}{!}{
% \begin{tabular}{|l|c|c|c|c|c|}
% \hline
% Loss Function    & Acc$_7^\uparrow$  & Acc$_2^\uparrow$  & F1$^\uparrow$    & MAE$^\downarrow$   & Corr$^\uparrow$  \\ \hline
% L1 & 50.9 & 79.7 & 80.1 & 0.639 & 0.631 \\
% L2            &   {51.1}     &   {80.0}  &   {80.3}  &   {0.635} &   {0.637} \\
% \bottomrule

% \end{tabular}
% }
% \label{Table:loss-abalation}
% \vspace{-3mm}
% \end{table}

\begin{table}[t]
\centering
\small
\caption{The effectiveness of different consistency loss. The results are reported on the CMU-MOSEI aligned dataset~\cite{cmu-mosei}.}
\resizebox{0.9\linewidth}{!}
{
\begin{tabular}{c|c|c|c|c|c}
\toprule
Loss Function    & Acc$_7^\uparrow$  & Acc$_2^\uparrow$  & F1$^\uparrow$    & MAE$^\downarrow$   & Corr$^\uparrow$  \\ \hline
%\vspace{-1em}
L1 & 50.9 & 79.7 & 80.1 & 0.639 & 0.631 \\
L2            &   {51.1}     &   {80.0}  &   {80.3}  &   {0.635} &   {0.637} \\
\bottomrule

\end{tabular}
\label{Table:loss-abalation}
}
\end{table}

\subsection{Ablation Study}
In this section, we evaluate the effectiveness of each individual cross-modal transformer on the CMU-MOSEI dataset. Specifically, we analyze the performance when only utilizing the cross-modal attention module for the language modality, represented as $A' \rightarrow L$ and $V \rightarrow L$, the blue sub-branch in Fig.~\ref{Figure: pipeline}. We also conduct similar evaluations for the visual and acoustic modalities. As shown in Table~\ref{Table:abalation}, the highest performance is achieved when the target modality is text (language). Combining all the cross-modal transformers further improves the performance. Additionally, Table~\ref{Table:loss-abalation} shows the results of using two different consistency losses to constrain the reconstructed features on the CMU-MOSEI dataset. We find that using the L2 loss leads to better performance.

\section{Limitation, Future Work and Conclusion}
While our method demonstrates strong robustness under missing-audio conditions, several limitations remain. First, extending the framework to scenarios where the visual or textual modalities are absent is non-trivial and warrants future investigation. Second, although the reconstruction enhances predictive performance, it does not guarantee interpretability or perceptual fidelity of the generated audio features. Finally, the reliance on transformer-based architectures introduces higher computational overhead compared to lightweight fusion models, which may restrict deployment in real-time or resource-constrained environments.

In conclusion, we present a knowledge-transfer network which utilizes a consistency loss for the task of multimodal learning with a missing modality for multimodal sentiment analysis. Our method specifically reconstructs the missing modal information and fuses it with the available modalities through a cross-modal attention network. Through extensive experiments on three sentiment analysis benchmarks, we demonstrate that the proposed method outperforms other baseline approaches and is capable of achieving comparable results to fully supervised multi-modality methods.




\printbibliography

@inproceedings{tsai2019multimodal,
  title={Multimodal transformer for unaligned multimodal language sequences},
  author={Tsai, Yao-Hung Hubert and Bai, Shaojie and Liang, Paul Pu and Kolter, J Zico and Morency, Louis-Philippe and Salakhutdinov, Ruslan},
  booktitle={Proceedings of the conference. Association for Computational Linguistics. Meeting},
  volume={2019},
  pages={6558},
  year={2019},
  organization={NIH Public Access}
}

@article{cmu-mosi,
  title={Multimodal sentiment intensity analysis in videos: Facial gestures and verbal messages},
  author={Zadeh, Amir and Zellers, Rowan and Pincus, Eli and Morency, Louis-Philippe},
  journal={IEEE Intelligent Systems},
  volume={31},
  number={6},
  pages={82--88},
  year={2016},
  publisher={IEEE}
}

@inproceedings{cmu-mosei,
  title={Multimodal language analysis in the wild: Cmu-mosei dataset and interpretable dynamic fusion graph},
  author={Zadeh, AmirAli Bagher and Liang, Paul Pu and Poria, Soujanya and Cambria, Erik and Morency, Louis-Philippe},
  booktitle={Proceedings of the 56th Annual Meeting of the Association for Computational Linguistics (Volume 1: Long Papers)},
  pages={2236--2246},
  year={2018}
}

@article{iemocap,
  title={IEMOCAP: Interactive emotional dyadic motion capture database},
  author={Busso, Carlos and Bulut, Murtaza and Lee, Chi-Chun and Kazemzadeh, Abe and Mower, Emily and Kim, Samuel and Chang, Jeannette N and Lee, Sungbok and Narayanan, Shrikanth S},
  journal={Language resources and evaluation},
  volume={42},
  number={4},
  pages={335--359},
  year={2008},
  publisher={Springer}
}

@article{transformer,
  title={Attention is all you need},
  author={Vaswani, Ashish and Shazeer, Noam and Parmar, Niki and Uszkoreit, Jakob and Jones, Llion and Gomez, Aidan N and Kaiser, {\L}ukasz and Polosukhin, Illia},
  journal={Advances in neural information processing systems},
  volume={30},
  year={2017}
}

@inproceedings{pham2019found,
  title={Found in translation: Learning robust joint representations by cyclic translations between modalities},
  author={Pham, Hai and Liang, Paul Pu and Manzini, Thomas and Morency, Louis-Philippe and P{\'o}czos, Barnab{\'a}s},
  booktitle={Proceedings of the AAAI Conference on Artificial Intelligence},
  volume={33},
  pages={6892--6899},
  year={2019}
}

@article{tsai2018learning,
  title={Learning factorized multimodal representations},
  author={Tsai, Yao-Hung Hubert and Liang, Paul Pu and Zadeh, Amir and Morency, Louis-Philippe and Salakhutdinov, Ruslan},
  journal={arXiv preprint arXiv:1806.06176},
  year={2018}
}

@article{liang2018multimodal,
  title={Multimodal language analysis with recurrent multistage fusion},
  author={Liang, Paul Pu and Liu, Ziyin and Zadeh, Amir and Morency, Louis-Philippe},
  journal={arXiv preprint arXiv:1808.03920},
  year={2018}
}

@inproceedings{zadeh2018memory,
  title={Memory fusion network for multi-view sequential learning},
  author={Zadeh, Amir and Liang, Paul Pu and Mazumder, Navonil and Poria, Soujanya and Cambria, Erik and Morency, Louis-Philippe},
  booktitle={Proceedings of the AAAI conference on artificial intelligence},
  volume={32},
  year={2018}
}

@inproceedings{poria2017context,
  title={Context-dependent sentiment analysis in user-generated videos},
  author={Poria, Soujanya and Cambria, Erik and Hazarika, Devamanyu and Majumder, Navonil and Zadeh, Amir and Morency, Louis-Philippe},
  booktitle={Proceedings of the 55th annual meeting of the association for computational linguistics (volume 1: Long papers)},
  pages={873--883},
  year={2017}
}

@inproceedings{wang2019words,
  title={Words can shift: Dynamically adjusting word representations using nonverbal behaviors},
  author={Wang, Yansen and Shen, Ying and Liu, Zhun and Liang, Paul Pu and Zadeh, Amir and Morency, Louis-Philippe},
  booktitle={Proceedings of the AAAI Conference on Artificial Intelligence},
  volume={33},
  pages={7216--7223},
  year={2019}
}

@article{chen2018best,
  title={The best of both worlds: Combining recent advances in neural machine translation},
  author={Chen, Mia Xu and Firat, Orhan and Bapna, Ankur and Johnson, Melvin and Macherey, Wolfgang and Foster, George and Jones, Llion and Parmar, Niki and Schuster, Mike and Chen, Zhifeng and others},
  journal={arXiv preprint arXiv:1804.09849},
  year={2018}
}

@article{lee2018pre,
  title={Pre-training of deep bidirectional transformers for language understanding},
  author={Lee, J Devlin M Chang K and Toutanova, K},
  journal={arXiv preprint arXiv:1810.04805},
  year={2018}
}

@inproceedings{ctc,
  title={Connectionist temporal classification: labelling unsegmented sequence data with recurrent neural networks},
  author={Graves, Alex and Fern{\'a}ndez, Santiago and Gomez, Faustino and Schmidhuber, J{\"u}rgen},
  booktitle={Proceedings of the 23rd international conference on Machine learning},
  pages={369--376},
  year={2006}
}

@inproceedings{DBLP:conf/cvpr/ZhangGWZLCCRHYC16,
  author    = {Zheng Zhang and
               Jeffrey M. Girard and
               Yue Wu and
               Xing Zhang and
               Peng Liu and
               Umur A. Ciftci and
               Shaun J. Canavan and
               Michael Reale and
               Andrew Horowitz and
               Huiyuan Yang and
               Jeffrey F. Cohn and
               Qiang Ji and
               Lijun Yin},
  title     = {Multimodal Spontaneous Emotion Corpus for Human Behavior Analysis},
  booktitle = {CVPP},
  pages     = {3438--3446},
  year      = {2016}
}

@inproceedings{DBLP:conf/cvpr/ZengTPLZZHL05,
  author    = {Zhihong Zeng and
               Jilin Tu and
               Brian Pianfetti and
               Ming Liu and
               Tong Zhang and
               ZhenQiu Zhang and
               Thomas S. Huang and
               Stephen E. Levinson},
  title     = {Audio-Visual Affect Recognition through Multi-Stream Fused {HMM} for
               {HCI}},
  booktitle = {CVPR},
  pages     = {967--972},
  year      = {2005}
}

@inproceedings{DBLP:conf/iccv/GanWHJ17,
  author    = {Quan Gan and
               Shangfei Wang and
               Longfei Hao and
               Qiang Ji},
  title     = {A Multimodal Deep Regression Bayesian Network for Affective Video
               Content Analyses},
  booktitle = {ICCV},
  pages     = {5123--5132},
  year      = {2017}
}

@inproceedings{liu2020crnet,
  title={Crnet: Cross-reference networks for few-shot segmentation},
  author={Liu, Weide and Zhang, Chi and Lin, Guosheng and Liu, Fayao},
  booktitle={Proceedings of the IEEE/CVF conference on computer vision and pattern recognition},
  pages={4165--4173},
  year={2020}
}

@article{liu2025modality,
  title={Modality-Aware Feature Matching: A Comprehensive Review of Single-and Cross-Modality Techniques},
  author={Liu, Weide and Zhou, Wei and Liu, Jun and Hu, Ping and Cheng, Jun and Han, Jungong and Lin, Weisi},
  journal={arXiv preprint arXiv:2507.22791},
  year={2025}
}

\end{document}